\documentclass[11pt,a4paper]{article}
\pdfoutput=1

\usepackage{jheppub}
\notoc

\usepackage{amsfonts,makecell,subfigure,multirow}

\newcommand\SM{SM}
\newcommand\BSM{BSM}
\newcommand{\ie}{i.e.~}
\newcommand{\eg}{e.g.~}

\newcommand\eq[1]{\begin{align}#1\end{align}}
\newcommand{\GeV}{\mathrm{GeV}}
\newcommand{\TeV}{\mathrm{TeV}}
\newcommand{\M}{\mathcal{M}}
\newcommand{\sh}{\hat{s}}

\title{\boldmath $Z'$ at the LHC: Interference and Finite Width Effects in Drell-Yan}

\author[a,b]{Elena Accomando}
\author[c]{Diego Becciolini}
\author[a,b]{Alexander Belyaev}
\author[a,b]{Stefano Moretti}
\author[b]{Claire Shepherd-Themistocleous}

\affiliation[a]{School of Physics \& Astronomy, University of Southampton, Highfeld, Southampton SO17 1BJ, UK}
\affiliation[b]{Particle Physics Department, Rutherford Appleton Laboratory, Chilton, Didcot, Oxon OX11 0QX, UK}
\affiliation[c]{{\rm CP}$^{\bf 3}${\rm-Origins} \& the Danish Institute for Advanced Study {\rm DIAS}, University of Southern Denmark, Campusvej 55, DK-5230 Odense M, Denmark}

\emailAdd{e.accomando@soton.ac.uk}
\emailAdd{becciolini@cp3.dias.sdu.dk}
\emailAdd{a.belyaev@soton.ac.uk}
\emailAdd{s.moretti@soton.ac.uk}
\emailAdd{claire.shepherd@stfc.ac.uk}

\date{\today}

\preprint{
{\raggedleft
CP3-Origins-2013-013 DNRF90\\
DIAS-2013-13\\
SHEP-13-02\\}
}

\keywords{
Hadronic Colliders, Phenomenological Models
}

\arxivnumber{1304.6700}
\abstract{
The interference effects between an extra neutral spin-1
$Z'$-boson and the Standard Model background in the Drell-Yan channel at the LHC
are studied in detail.
The final state with two oppositely charged leptons is considered.
The interference contribution to the new physics signal, currently not fully taken into account by experimental collaborations in $Z'$-searches and in the interpretation of the results, can be substantial.
It may affect limits or discovery prospects of $Z'$ at the LHC.
As the $Z'$-boson interference is model-dependent, a proper treatment would a priori require a dedicated experimental analysis for each particular model.
Doing so could potentially improve the sensitivity to new physics, but would require significantly more experimental effort.

At the same time, it is shown that one can define an invariant mass window, valid for a wide range of models, for which the contribution of the model-dependent interference to the Beyond the Standard Model signal is reduced to $\mathcal{O}(10\%)$, % level,
comparable to the level of the combined uncertainty 
from parton densities and higher order corrections.
This quasi-model-independent ``magic cut" does not scale with the mass of the $Z'$-boson and is approximately constant over a large range of masses.
Such control of the interference effects relies on not-too-small branching ratios of $Z'$ to leptons (typically of at least a few percent) which can be suppressed, however, by additional new decay channels of the $Z'$; %, increasing the interference effect;
a small width-to-mass ratio alone %, on the other hand,
does \emph{not} guarantee the interference to be small over an arbitrary kinematic range.
Under the general  assumption that these new decay channels of $Z'$ 
are not dominant,
one can perform quasi-model-independent analyses, preserving the current scheme used by the experimental collaborations for the $Z'$-boson search using the
suggested invariant mass window cut.
}

\begin{document}

\maketitle
\newpage
%\tableofcontents

\section{Introduction}

Drell-Yan (DY) processes have been studied for over 40 years.
They involve the production of lepton-anti-lepton pairs with high invariant mass in hadron collisions~\cite{Drell:1970wh}.
They are of great historical importance, as the discoveries of the Standard Model (\SM) $W$ and $Z$ resonances at the Super Proton Synchrotron (SPS) in 1983 were indeed made in these channels~\cite{Arnison:1983rp, Banner:1983jy, Arnison:1983mk, Bagnaia:1983zx}.
At the same time, DY processes are very powerful for discovering or limiting new physics involving
heavier particles similar to the $W$ and $Z$ bosons.

Scenarios involving such exotic particles are particularly relevant now as the Large Hadron Collider (LHC) is collecting data and probing higher and higher energies.
The signals of the Drell-Yan channels are among the cleanest
in the difficult environment of a high-energy hadron collider.
The final state one tries to identify is purely leptonic and stands out from the considerable Quantum Chromodynamical (QCD) activity.
No colour charge in the final state implies fewer diagrams, making it a simple process to study.
On the other hand, it also means that the cross-section is lower compared to other (non-leptonic) processes.
Less background and easier identification ensure a high efficiency, though, which
compensates for the reduced event-rate.

The contribution of new physics to any given process is, in general, not independent of the known physics: there may be interference between the two, and it can be sizeable.
This happens in particular in the Drell-Yan channels with additional resonances.
Although the effect of interference is commonly discussed in this specific context (or in closely related ones)~\cite{Dittmar:1996my, Rizzo:2007xs, Petriello:2008zr, Papaefstathiou:2009sr, Rizzo:2009pu, Chiang:2011kq, Accomando:2011eu, Choudhury:2011cg, Leike:1998wr},
one still encounters misleading or incomplete statements regarding, for instance, how the importance of interference depends on kinematic cuts and other parameters and how large the effect is.

Up until now, most experimental searches of new particles in these channels do not explicitly include interference effects when interpreting the available data~\cite{Aaltonen:2010jj, Aaltonen:2008ah, Abazov:2007ah, Abazov:2010ti, Aad:2012dm,
ATLAS-CONF-2013-017, Chatrchyan:2013lga, Chatrchyan:2012oaa}.
There is, however, a growing interest in including this effect in the community, and some of the more recent analyses do discuss the matter~\cite{Chatrchyan:2012meb, Aad:2012hf}, at least when considering cases where interference has explicitly been shown to be important~\cite{Bella:2010sc}.
Interestingly, the effect of interference in a different but related channel (top-bottom quark pair~\cite{Boos:2006xe}) has been included in analyses for some time now~\cite{Abazov:2008vj, Abazov:2011xs, Chatrchyan:2012gqa}.

It is therefore important to study in detail and illustrate certain features of the interference between the \SM\ electroweak bosons and potential new resonances in DY processes, namely the relative size of the interference and main factors which affect it.

The focus of this discussion is on the neutral channel at the LHC, \ie the production of an electron or muon pair mediated by a photon, a $Z$ and a hypothetical $Z'$-boson in proton-proton collisions.
$Z'$ here denotes an extra neutral spin-1 particle.
No flavour changing effects are considered.

First, a comparison with the Narrow Width Approximation (NWA) is presented, in which production and decay of the intermediate particle are separated.
This exemplifies in the current context one of the limitations that have been pointed out~\cite{Berdine:2007uv, Kauer:2007zc, Uhlemann:2008pm}.
Some previous statements are also reviewed~\cite{Accomando:2010fz}.
The general trend is shown for a large number of typical $Z'$ models.
The importance of these observations is that properties of $Z'$ bosons are often expressed by means of the so-called $c_u$-$c_d$ parametrisation, which implicitly relies on the NWA~\cite{Carena:2004xs}.

However, of perhaps greater phenomenological relevance is a comparison not with the NWA result, but with a Finite Width computation neglecting interference.
One should determine how far away from the resonance region it is still safe to neglect the interference.
The dependence of the effect on the mass and width of the new resonance is
discussed.
Some more specific results are given in the context of the most common (but not particularly well theoretically motivated) benchmark scenario, the Sequential Standard Model (SSM), where the $Z'$ is assumed to have the same fermion couplings as the \SM\ Z (see for instance ref.~\cite{Altarelli:1989ff} and relevant experimental analyses),
and of the so-called $\psi$ model, motived by $E_6$ Grand Unified Theories (GUT).
The paper is organised as follows:
the general framework is defined in Section II,
followed by Section III where benchmark models are introduced, and Section IV with computational details.
Section V and VI discuss results on finite width effects
and size of the interference followed by conclusions in Section VII.

\section{Interference}

\subsection{In general}

Interference generically refers to the cross-term (in the probability of a given process) between two different contributions to the transition amplitude.
It is typical of wave and quantum mechanics.
In particle physics, the amplitude is essentially given by a matrix element $\M = \sum_i \M_i$ which corresponds, in perturbation theory, to a sum of individual Feynman diagrams.
The interference is then the second sum of
\eq{
{\bigl| \M \bigr|}^2 = \sum_i {\bigl| \M_i \bigr|}^2 + \sum_{i<j} 2 \operatorname{Re}\left(\M^*_i \M_j\right),
}
while the first terms are the ``diagonal'', or ``pure'', contributions.

The relative size of the interference in a particular process (\ie considering a single ``entry'' of the ``matrix'' $\M$) only depends on the relative size (and phase) of the different $\M_i$.
In practice, however, one is never interested in a single specific transition:
different initial and final states are summed or averaged over (momenta, helicities, polarizations, colour charges, etc.).
It may happen that two different contributions are only sometimes, or even never, simultaneously non-zero in the set of considered sub-processes and therefore do not interfere with each other.
Thus, the interference may not be maximal, or may completely vanish in very specific cases.
For instance, this is the case when considering particles with opposite helicity structure in their interactions;
or the interference between a scalar and a gauge boson is typically suppressed by the masses of the external particles.

What determines the presence or absence of interference, as well as its size and
sign,\footnote{Indeed, contrary to the diagonal terms ${| \M_i |}^2$, interference is not positive-definite.} can be --- somewhat arbitrarily --- separated into two categories.

The first factor is determined by the  model details;
in particular the coupling structure of the different particles considered here (\ie what precise states they couple to), as this is what decides whether there is interference or not to begin with.
In that sense, interference is highly model dependent.
Then, for a given interaction structure, the overall size of couplings should a priori fix the relative importance of the effect;\footnote{The width dependence of the results makes the picture a little more complicated, though.}
masses essentially set the kinematic dependence.

The second ingredient affecting the interference is kinematics:
the different contributions to the amplitude will in general not all have the same dependence on kinematic variables.
In fact, kinematic variables of interest are precisely constructed in order to have separate regions where one contribution or the other dominates the process (\eg a new-physics-free region and a low \SM\ background region).
This implies that, somewhere between two such regions (provided the kinematic dependence is smooth), these contributions need to be of the same importance;
thus, the interference between them would \emph{not} be kinematically suppressed at this point.
Therefore, unless there is a strong overall suppression because of a particular interaction structure, there has to be a kinematic region where the interference is important.
One of the most representative  of such kinematical variables is the di-lepton invariant mass $M_{\ell\ell}$:
the region roughly half-way between the $Z'$ and the \SM\ resonances would for instance be the ``intermediate region"
mentioned above where the interference is important.
Finally, one should note that outside of this intermediate kinematic range, interference should be larger than the sub-dominant diagonal contribution.
In our particular example above, in the vicinity of the \SM\ resonances, the pure Z' contribution is less than the interference;
conversely, the interference becomes larger than the \SM\ background near the heavy resonance.

\subsection{In Drell-Yan}

The process under consideration is the production of a charged lepton pair at the LHC in the neutral Drell-Yan channel:
\eq{
pp\rightarrow \gamma , Z, Z' \rightarrow l^+l^-,
}
which is mediated by the \SM\ photon and $Z$-boson, and possibly by an extra $Z'$-boson.
As all results are given at parton level, the conclusions of this article apply to both electron and muon pairs.
Notice that, even without new physics beyond the \SM\ (\BSM), there is interference between the photon and the $Z$-boson.
Although --- as will be discussed --- the effect is very suppressed, observation of asymmetries largely due to this interference even allowed a glimpse of the $Z$-boson in certain channels before the resonance peak itself was accessible~\cite{Prescott:1978tm, Wu:1984ik}.

\begin{figure}[tbp]
\centering
\subfiguretopcaptrue
\subfigure[]{
\includegraphics[width=0.47\textwidth]{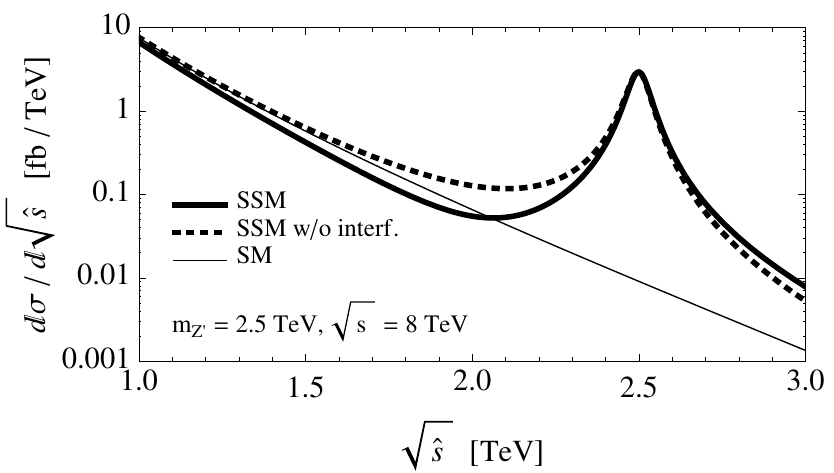}
\label{fig:distrib}
}
\subfigure[]{
\includegraphics[width=0.47\textwidth]{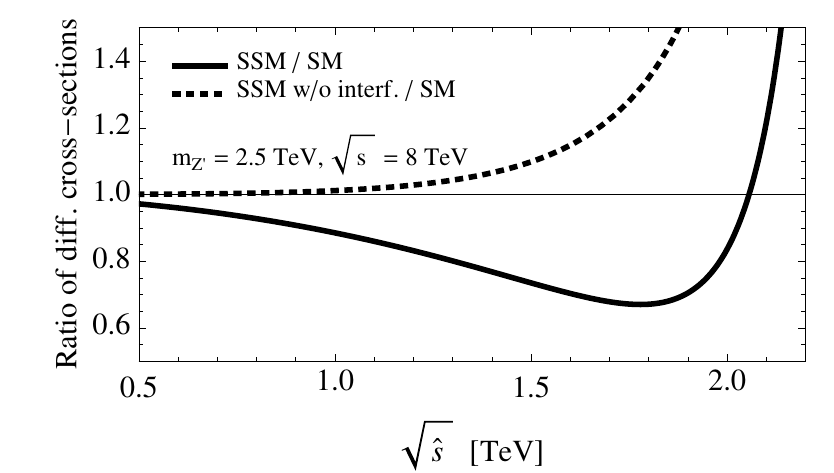}
\label{fig:ratios}
}
\caption[Interf. in DY]{
\subref{fig:distrib} Invariant mass distribution in the presence of a SSM $Z'$ resonance, computed with and without interference between the $Z'$ and the \SM. \subref{fig:ratios} Ratio of these two predictions to the pure \SM\ Drell-Yan result. LHC@8TeV with $M_{Z'}$=2.5 TeV is considered. No kinematical cuts are applied and the CT10 NLO best fit PDF set is used.
}
\label{fig:interf_DY}
\end{figure}

There are some prospects of measuring such effects due to an extra $Z'$-boson at the LHC~\cite{Dittmar:1996my}, but, in popular scenarios at least, it is unlikely to allow a discovery before the observation of the expected peak in the invariant mass distribution~\cite{Rizzo:2009pu}.
Nevertheless, interference may be significant in the intermediate invariant mass range, as shown in figure~\ref{fig:interf_DY}.
Indeed, one can see that
the effect of interference in the differential
cross-section
can be as large as $50\%$.

The current discussion is concentrated on the cross-section in the peak region which is the focus of current search strategies.

Neglecting the masses of external fermions, if the scattering angle is integrated over, and no asymmetric acceptance cut is used (\ie different rapidity cut for particle and anti-particle), parity-odd effects drop out and the only relevant coupling factors appearing in the matrix element squared are
\eq{
\M^*_i \M_j \propto
\left( g_V^i g_V^j + g_A^i g_A^j \right)_\text{quark} \left( g_V^i g_V^j + g_A^i g_A^j \right)_\text{lepton},
\label{eq:coupling}
}
where $g_V^i$ and $g_A^i$ are the vector and axial couplings describing the neutral current interaction
\eq{
\mathcal{L}_{NC} = Z_\mu^i\ \bar{\psi}_f\, \gamma^\mu \left( {g_V^i}_f - {g_A^i}_f\, \gamma_5 \right) \psi_f;
}
$i$ labels the neutral gauge boson (photon, Z, $Z'$) and $f$ the fermion type and flavour.

The kinematic dependence is given by the propagator factors
\eq{
\frac{\left( \sh - m_i^2 \right) \left( \sh - m_j^2 \right) + \left( m_i\, \Gamma_i \right) \left( m_j\, \Gamma_j \right)}{\left( \left(\sh - m_i^2\right)^2 + m_i^2\, \Gamma_i^2 \right) \left( \left(\sh - m_j^2\right)^2 + m_j^2\, \Gamma_j^2 \right)},
\label{eq:propag}
}
where $\sh$ is the centre-of-mass energy of the partonic process squared; $m_i$ and $\Gamma_i$ are the mass and width of the resonance $i$.

From now on, \emph{interference} will be used to refer to the interference between a hypothetical $Z'$ (the new physics) and the \SM\ photon and $Z$-boson (the known physics).

Whether the interference is constructive or destructive depends on the relative sign of \eqref{eq:coupling} and \eqref{eq:propag}.
The propagator factor in interference contributions, as a function of $\sh$, always changes sign when crossing both resonance peaks; in the region of interest --- between the two resonances --- it is negative.
If the coupling factor is positive, in particular if the couplings are sequential, the interference is destructive in the intermediate range.

Finally, interference terms contain two powers of the coupling to each relevant resonance, and diagonal terms four powers of the coupling; this thus determines the scaling of the different contributions when varying the overall size of the couplings. A $Z'$-boson with large couplings to fermions compared to the \SM\ gauge bosons will exhibit a smaller relative interference, and vice-versa.

\section{\boldmath Benchmark models with an additional $U'(1)$ group}

The prediction of (at least) one extra neutral vector boson is a generic feature of models where the gauge group is extended compared to the \SM.
Scenarios falling in that category include Grand Unified Theories (GUT), theories of dynamical electroweak symmetry breaking and extra-dimensional theories (see~\cite{Langacker:2008yv} and references therein).
The set of models that are being specifically considered here is the same as in~\cite{Accomando:2010fz}.
It consists of thirteen different models, split in three classes:
GUT with an $E_6$ gauge group ($E_6$), Generalised Left-Right symmetric models (GLR), and generalisations of the SSM benchmark scenario (GSM).

No flavour changing effects are considered, and family universality of the fermion couplings is assumed.
The $Z'$ resonances are taken to be as narrow as allowed by the commonly adopted assumption that only the direct decay to \SM\ fermions is allowed.
The widths in these models are typically a few percent of the $Z'$-boson mass, and the branching ratios (BR) to charged leptons of a given flavour fall in a similar range, as shown in table~\ref{tab:Zfactors}.
An exception is given by the $Q$-model (last row in table~\ref{tab:Zfactors})
where the values of these observables exceed the $\mathcal{O}(10\%)$ level.
The $Q$-model is here taken as representative of wide vector resonances.    
Because the photon and the $Z$-boson have different chiral interaction structures, a single $Z'$ cannot interfere maximally with both
(nor can the interference identically vanish for any choice of couplings);
therefore, there will always be some suppression of the effect in these models, unlike in cases where there are \emph{two} extra (degenerate) resonances~\cite{Bella:2010sc}.
As will be shown, this suppression is still not enough to guarantee the interference to be negligible.

\begin{table}[tbp]
\small
\centering
$
\begin{array}{|cc|rrr|rrr||r|r|}
 \multicolumn{2}{c|}{}  & \multicolumn{3}{c|}{\text{down-type quark}} & \multicolumn{3}{c||}{\text{up-type quark}} & \Gamma/m [\%] & \text{BR} [\%] \\
\hline\hline
 \multicolumn{2}{|c|}{\diaghead(3,-1){------------------}{$\quad i \quad$}{$\quad j \quad$}} & \gamma  & Z & Z' & \gamma  & Z & Z' & & \\
\hline
 & \gamma  & \mathbf{1} &  &  & 4 &  &  & & \\
 & Z & 0.06 & 1.66 &  & 0.06 & 1.28 &  & 2.7 & 3.4 \\
 \hline
 \multirow{6}{*}{$E_6$} & \chi & -0.65 & -0.08 & 0.66 & 0 & 0.20 & 0.13 & 1.2 & 6.1 \\
 & \psi  & 0 & 0.29 & 0.07 & 0 & -0.29 & 0.07 & 0.5 & 4.4 \\
 & \eta  & -0.24 & 0.07 & 0.07 & 0 & -0.09 & 0.12 & 0.6 & 3.7 \\
 & S & -0.55 & 0.04 & 0.78 & 0 & 0.11 & 0.02 & 1.2 & 6.6 \\
 & I & -0.41 & 0.14 & 0.66 & 0 & 0 & 0 & 1.1 & 6.7 \\
 & N & -0.04 & 0.30 & 0.16 & 0 & -0.26 & 0.07 & 0.6 & 5.6 \\
 \hline
 \multirow{4}{*}{GLR} & R & 0.67 & 0.20 & 1.81 & 1.34 & 0.40 & 1.81 & 2.4 & 4.8 \\
 & B-L & -0.90 & -0.05 & 0.80 & 1.79 & 0.03 & 0.80 & 1.5 & 15.4\\
 & LR & 0.10 & 0.07 & 0.87 & 0.11 & 0.43 & 0.50 & 2.0 & 2.4 \\
 & Y & 0.34 & 0.21 & 1.25 & 3.36 & 0.10 & 4.27 & 2.3 & 12.5 \\
 \hline
 \multirow{3}{*}{GSM} & SSM & 0.06 & 1.66 & 1.66 & 0.06 & 1.28 & 1.28 & 3.0 & 3.1 \\
 & T_{3L} & 1.11 & 2.13 & 4.94 & 2.22 & 1.74 & 4.94 & 4.6 & 4.2 \\
 & Q & 5.93 & 0.33 & 35.1 & 23.7 & 0.36 & 141. & 12.3 & 12.5 \\
 \hline
\end{array}
$
\caption[Couplings factors in $Z'$ models]{Value of the factor \eqref{eq:coupling} for all contributions in all considered models, normalised to the photon coupling in down-quark channels ($e^4 / 9$).
The last two columns give the widths and branching ratios to charged leptons of the $Z'$-boson.}
\label{tab:Zfactors}
\end{table}

In most $E_6$ models, the interference contributions in the processes with up- and down-type quarks in the initial state have opposite signs, thus they partly cancel each other, resulting in reduced interference in the full Drell-Yan process. The overall interference is also generally constructive in the intermediate region, unlike in the other cases.
A similar cancellation occurs for some values of the mixing angle in the LR models.
Since the relative importance of the up- and down-type quark channels --- in other words the ratio of parton luminosities --- varies with the momentum fractions, these cancellations depend on the partonic centre-of-mass energy, and additional sign changes of the interference can occur.

These features can be read off table~\ref{tab:Zfactors}, which shows the value of all the relevant coupling factors \eqref{eq:coupling}.
For instance in the $E_6$ $\psi$ model, there is no interference between the $Z'$ and the photon (as the former is an axial resonance), but the interference factors with the $Z$ are large compared to the pure $Z'$ couplings. They have opposite signs in the two different quark channels, however, which leads to a suppression of the overall interference. Furthermore since the up-quark parton luminosity is larger in the considered context, the negative coupling factor will dominate, leading to constructive interference below the $Z'$ peak.

Also note the $\mathcal{O}(10^{-2})$ suppression of interference in the \SM\ between the photon and the Z.

\section{Remarks on the computation}

The observable considered here is the cross-section $\sigma (pp\rightarrow l^+l^-)$ for producing the leptons within a certain invariant mass window symmetric around the $Z'$ mass,
\eq{
\bigl| m_{ll} - m_{Z'} \bigr| < \Delta m,
\label{eq:window}
}
and results are presented as a function of $\Delta m$, the cut on the dilepton invariant mass, or more precisely in terms of $\Delta m / \sqrt{s}$.
Indeed, the main effect of varying the hadronic centre-of-mass energy $\sqrt{s}$ between $7$ and $14\,\TeV$ here is related to a simple rescaling of $\Delta m$ and $m_{Z'}$:
all the following figures are essentially valid for any LHC energy.
Computations have all been performed choosing $\sqrt{s} = 8\,\TeV$.

Because of the very steep fall-off of the parton distribution functions (PDF) with the partonic centre-of-mass energy, where the upper bound of the integration range lies is of little relevance;
what essentially matters in \eqref{eq:window} is the lower bound on the lepton pair invariant mass.

As to numerical details, the following input parameters were chosen:
$m_Z = 91.1875\ \GeV$, $\alpha_{em}^{-1}=128.88$,  $\sin\theta_W^2=0.2304$, $G_F=1.1663787\times 10^{-5}\  \GeV^{-2}$.

All results are presented at parton level and at leading order (LO). Meaning that, in the context of this discussion, the invariant mass of the outgoing lepton pair $m_{ll}$ and the (partonic) centre-of-mass energy $\sqrt{\sh}$ are equal.
Next-to-leading order (NLO) QCD corrections are expected to approximately factor out from the process, as only the initial state carries colour charge and it is the same for all the diagrams.
The effect would be a general enhancement of the contributions.
The current discussion focuses on relative sizes, however, not on the absolute value of the prediction; the conclusions should therefore be robust against k-factors.
This also applies to acceptance cuts (typically requiring the leptons to have a pseudo-rapidity less than about 2.5): none have been implemented, but as the events are concentrated in the central region, including them would have little effect on the overall results.

Finally, although the hard scattering process is computed at LO, it is convoluted with the CT10 NLO best fit PDF set~\cite{Lai:2010vv} (choosing $\sqrt{\sh}$ for the factorisation scale).\footnote{This seemingly inconsistent procedure may in fact give a closer approximation of the full NLO result~\cite{Campbell:2006wx}.}

All computations have been performed with Mathematica~\cite{Mathematica7} and cross-checked with CalcHEP~\cite{Belyaev:2012qa}.

\section{Finite Width effects}

The NWA is a convenient approximation, both computationally and conceptually:
unstable particles are considered to be produced on-shell exclusively, allowing a factorisation of the process into a production rate and a BR (\ie the probability that the unstable particle decays into the desired final state). Computational power has increased considerably with time; in many cases, the complexity of a full computation is not so much an issue any more. On the other hand, many concepts derived from the NWA are still very much in use when it comes to describing new physics: production cross-sections and BR are widely used notions.
Unfortunately, the NWA can only be trusted under certain conditions and is often assumed to work better than it does~\cite{Berdine:2007uv, Kauer:2007zc, Uhlemann:2008pm}.

The coupling properties of $Z'$ bosons are commonly described in terms of two coefficients, $c_u$ and $c_d$~\cite{Carena:2004xs}.
They are simply the quantity \eqref{eq:coupling} for the diagonal $Z'$ contributions to the up- and down-type quark channels, up to a factor $\Gamma_{Z'}/m_{Z'}$ and some numerical coefficient.
The difference between a computation including off-shell effects (but not interference) and the NWA is just in how the propagator factor and the parton luminosities are treated, which only indirectly depend on the couplings (through the width of the resonance).
The functional form of the kinematic dependence of the up- and down-quark contributions being similar, one can find an almost model-independent invariant mass integration range such that the NWA and the Finite Width results agree.\footnote{Provided the width of the resonance is similar in all the models considered.}

The existence of such a ``magic'' cut was pointed out previously~\cite{Accomando:2010fz}.
A priori, one would expect this cut to depend on the mass of the resonance, but as the parton luminosities take a similar form over most of the $\sqrt{\sh}$ range, it turns out that it is the \emph{absolute} size of the magic integration window that varies little with the $Z'$ mass.
More precisely, $\sqrt{s}$, the hadronic centre-of-mass energy is the relevant scale in this discussion. In this paper, results are thus plotted as a function of $\Delta m / \sqrt{s}$ (rather than adopting the common convention $\Delta m / m_{Z'}$).

\begin{figure}[tbp]
\centering
\subfiguretopcaptrue
\subfigure[]{
\includegraphics[width=0.47\textwidth]{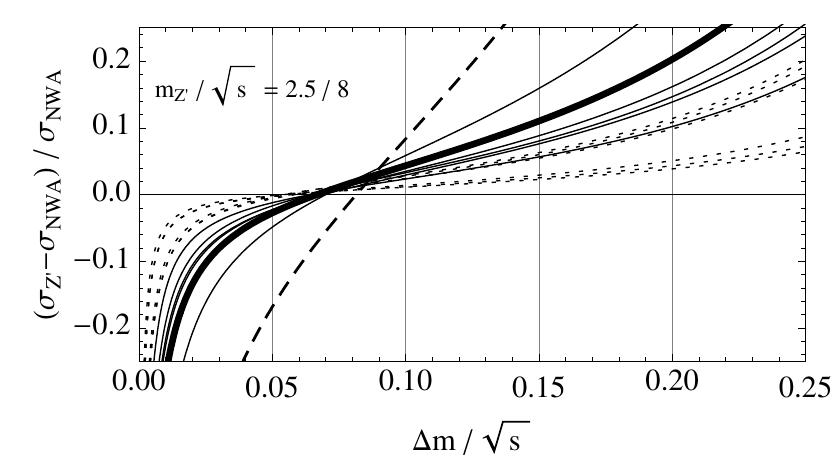}
\label{fig:nwa_without}
}
\subfigure[]{
\includegraphics[width=0.47\textwidth]{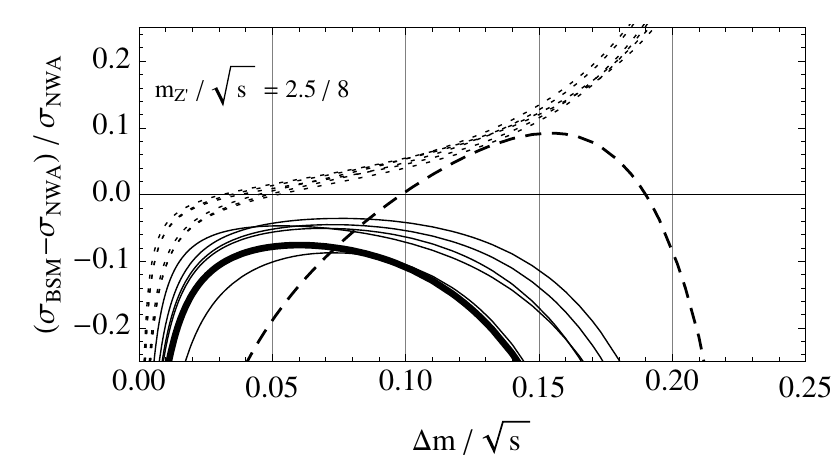}
\label{fig:nwa_with}
}
\caption[NWA vs no interf.]{
Relative difference between NWA and Finite Width computation \emph{excluding} \subref{fig:nwa_without} or \emph{including} \subref{fig:nwa_with} interference, as a function of $\Delta m / \sqrt{s}$, at a fixed $Z'$ mass.
All the models considered in~\cite{Accomando:2010fz} are plotted:
the thick line is the SSM result;
the dashed line is the $Q$-model;
the dotted lines are for the $E_6$ ones;
other models (GLR and $T_{3L}$) are thin continuous lines. LHC@8TeV with $M_{Z'}$=2.5 TeV is considered. No kinematical cuts are applied and the CT10 NLO best fit PDF set is used.
}
\label{fig:nwa}
\end{figure}

\begin{figure}[tbp]
\centering
\subfiguretopcaptrue
\subfigure[]{
\includegraphics[width=0.47\textwidth]{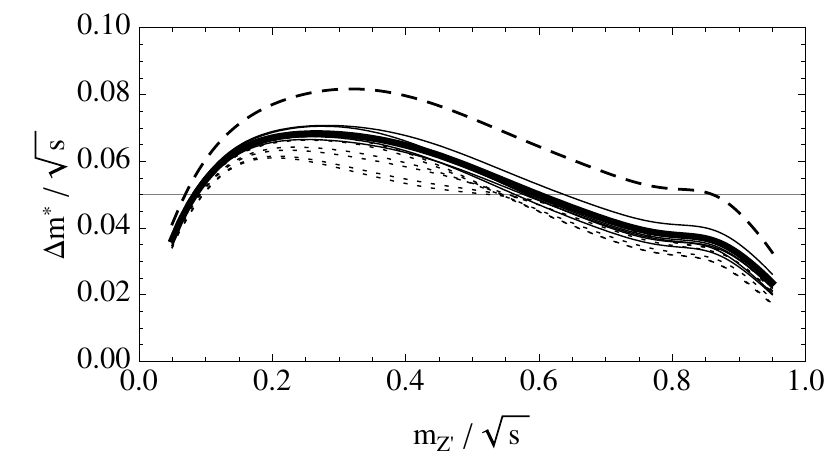}
\label{fig:mass_magic}
}
\subfigure[]{
\includegraphics[width=0.47\textwidth]{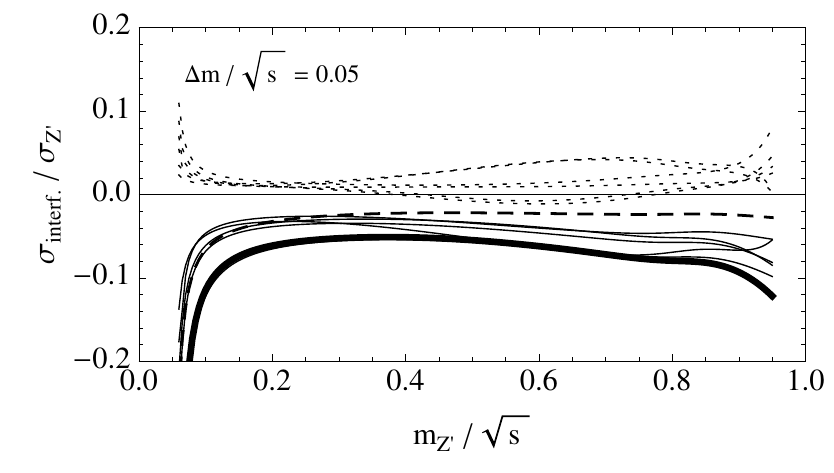}
\label{fig:mass_interf}
}
\caption[mass dependence]{
\subref{fig:mass_magic} Value of the cut $\Delta m^*/ \sqrt{s} $ for which NWA and the Finite Width computation without interference coincide in all the different models, as a function of $m_{Z'}/ \sqrt{s} $.
\subref{fig:mass_interf} Relative size of interference as a function of  $m_{Z'}/ \sqrt{s} $ for a fixed value of the cut $\Delta m / \sqrt{s} = 5 \%$.
The convention for the lines is the same as in figure~\ref{fig:nwa}.
}
\label{fig:mass}
\end{figure}

The agreement between the $Z'$-boson production cross-section computed in the NWA and calculated taking into account Finite Width effects, but neglecting interference, is reproduced in figure~\ref{fig:nwa_without}.
All the considered models predict for the $Z'$-boson a narrow width of $\mathcal{O}(1\%)$ of its mass, except the so-called $Q$-model, which has a relative width of order $10\%$ and is taken as a benchmark for wide vector resonances.
This latter model thus exhibits a slightly different behaviour.
In the comparison between NWA and Finite Width result with no interference, the important point is that the ``magic'' cut is quasi-model-independent, and its value is a little above $\Delta m / \sqrt{s} \sim 5 \%$.
Furthermore, if the agreement between these two approximations becomes more sensitive to the cut as the $Z'$-boson mass is increased, the approximate position of the ``magic cut" on the other hand varies little (figure~\ref{fig:mass_magic}).

Unfortunately, interference spoils this ideal picture, as shown in figure~\ref{fig:nwa_with}.
In the $E_6$ models, with small constructive interference, there is still an approximate ``magic cut", slightly shifted to lower $\Delta m$.
If the interference is destructive, it is impossible in general to exactly match the NWA cross-section and the value obtained from a full computation, no matter what invariant mass cut is used.
In all classes of models considered in the paper, one gets at best a $5$-$10\%$ overestimation when using the NWA.

A further comment is that changing the width by consistently scaling the couplings (\ie keeping the BR constant) or doing it while keeping the couplings relevant to the process constant (\ie changing the BR with constant partial width) is \emph{not} equivalent because of interference.
A good agreement with the NWA using a large integration window would be expected in the small width limit regardless, but it happens only if the limit is taken with constant couplings --- increasing the BR --- which is in principle inconsistent with the basic assumptions on the $Z'$ properties.\footnote{For the Drell-Yan process to happen in the first place, the $Z'$ needs to couple to at least one type of quark and one type of lepton, thus the branching ratio to the leptonic final state is necessarily less than one.}
If the width of the resonance is reduced by consistently making the couplings smaller, there will indeed be a better agreement between the full computation and the NWA, but only if a tighter invariant mass cut is applied;
if a sizeable off-shell region is included, the importance of the interference stays the same.
In other words, the interference is negligible when comparing the NWA and a full computation only if the width is narrow \emph{and} the BR large:
a small value of $\Gamma / m$ alone is \emph{not} a sufficient condition for the interference effects to be negligible (if one has to consider a larger kinematic region than the very narrow peak region, for instance because of a limited energy resolution).

Results illustrating these points are presented in the next section.

\section{The size of interference}

\begin{figure}[tbp]
\centering
\subfiguretopcaptrue
\subfigure[]{
\includegraphics[width=0.47\textwidth]{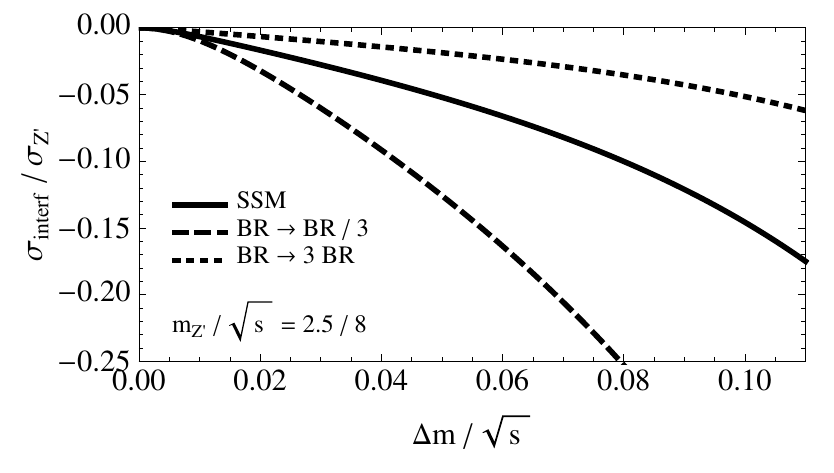}
\label{fig:widths_ssm}
}
\subfigure[]{
\includegraphics[width=0.47\textwidth]{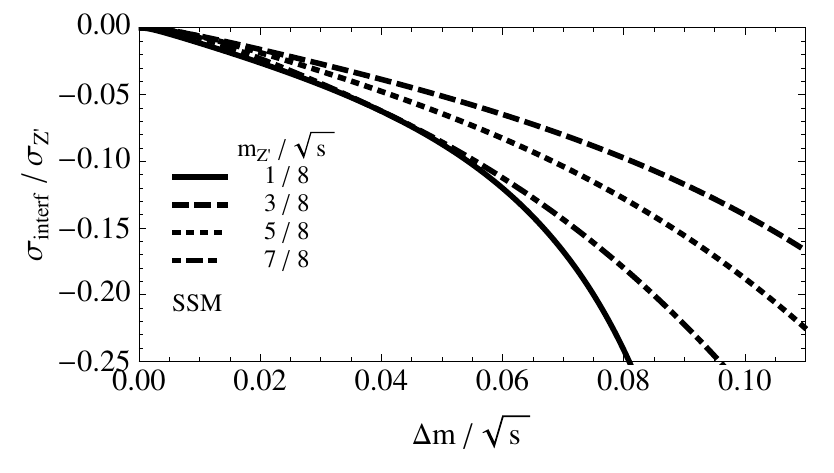}
\label{fig:masses_ssm}
}
\subfigure{
\includegraphics[width=0.47\textwidth]{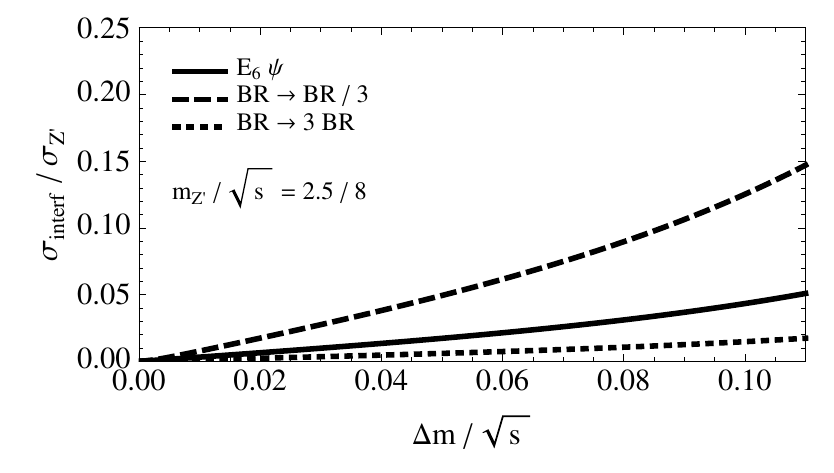}
\label{fig:widths_psi}
}
\subfigure{
\includegraphics[width=0.47\textwidth]{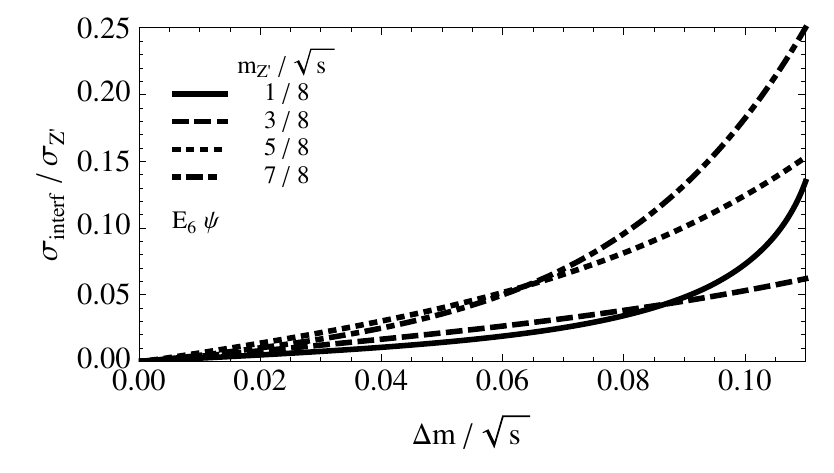}
\label{fig:masses_psi}
}
\caption[w/wo interf. vs masses/widths]{
Relative importance of the interference contribution to the \BSM\ signal (normalised to the pure $Z'$ cross-section), as a function of $\Delta m / \sqrt{s}$, varying the $Z'$ width (keeping couplings fixed) \subref{fig:widths_ssm} (left) or mass \subref{fig:masses_ssm} (right), in the SSM and in the $E_6$ $\psi$ model.
}
\label{fig:interf_ssm}
\end{figure}

Since the NWA is not so relied on any more in the context of the simple Drell-Yan processes, a comparison between the Finite Width predictions including or not interference is perhaps more interesting.
The difference between the two predicted cross-sections (\ie the interference contribution) relative to the pure $Z'$ result is plotted as a function of the same invariant mass window as before in figure~\ref{fig:interf_ssm}, for $Z'$-bosons of different masses and widths.
Two models are taken as sample, the SSM (upper plots) and the $E_6$ $\psi$ model (lower plots).

As can be seen, choosing an integration range as large --- or larger --- than prescribed by the magic matching between NWA and Finite Width results shown in figure~\ref{fig:mass_magic}, the importance of the interference to the \BSM\ cross-section increases to $\mathcal{O}(10\%)$, becoming comparable to many other corrective effects normally taken into account (NLO corrections, PDF uncertainties, experimental efficiencies).

In order to keep interference under control in all considered models, one should fix a small enough invariant mass cut, for instance the one prescribed by the matching between NWA and the Finite Width computation.
Choosing $\Delta m / \sqrt{s} \sim 5 \%$ or less, the interference does stay below about $10\%$ over a large range of masses (figure~\ref{fig:mass_interf}).
The important message here is that approximately model-independent statements can be made provided that the considered search window is sufficiently narrow.

This procedure has some limitations, though.
The main one being that interference noticeably grows in importance if the $Z'$-boson width is larger due to additional decay modes (for instance into the \SM\ gauge bosons).
This is shown in figure~\ref{fig:widths_ssm} for two representative models:
the SSM (upper plot) and the $E_6$ $\psi$ model (lower plot).
For a given cut, if the $Z'$ width is increased without changing the couplings (\ie the BR is decreased, the partial width staying constant), the interference contribution becomes larger, and vice-versa.\footnote{Once again, it is actually inconsistent to arbitrarily increase the BR.}
This also illustrates the last point of the previous section.
On the other hand, a change in the width done by consistently rescaling the couplings has much less impact on the importance of interference (and has an opposite effect) and is thus not shown here.

Figure~\ref{fig:masses_ssm} completes the discussion on the $m_{Z'}$ dependence by showing that the relative interference contribution as a function of the invariant mass cut does indeed not change dramatically with the $Z'$ mass, however only for a narrow enough search window.

Last, it should be pointed out that the complete \BSM\ signal, \ie pure $Z'$-boson production plus its interference with the \SM\ background without any kinematic cuts, can be very different from the NWA cross-section $\sigma (pp\rightarrow Z')\times \mathrm{BR}(l^+l^-)$.
This latter is the \mbox{(pseudo-)observable} popularly used when presenting the $95\%$ C.L. upper bound on the \BSM\ cross-section from which experimental bounds on the $Z'$-boson mass are extracted.
In most cases, the interference is generally negative;
thus, the \BSM\ contribution to the total cross-section is not necessarily positive and is not an appropriate physical quantity for expressing limits.
Furthermore, the physical meaning of such an observable is limited since, when a very large kinematic region is considered, the Drell-Yan cross-section is dominated by the \SM\ contribution, and interference, which then constitutes the major component of the \BSM\ signal, is drowned in the background.

\section{Summary and conclusion}

Hypothetical new physics is in general not independent from the known \SM:
there might be interference.
This happens in particular in Drell-Yan channels with extra resonances.
Such interference is often thought to be negligible regardless of the physical observable considered or of model details.

The results presented here show that even in the presence of a single $Z'$, where the interference cannot be maximal, it may still become an important effect when considering the \BSM\ cross-section in a large enough kinematic range.

The NWA estimate of the $Z'$ cross-section cannot in general represent the true prediction of a model, no matter what kinematic cut is chosen.
There is no ``magic'' invariant mass cut allowing a perfectly model-independent analysis.
However, the importance of interference --- and thus the deviation from this model-independent behaviour --- is limited if one does use the magic cut appearing when interference is neglected;
this cut is still a good measure of the typical range beyond which off-shell effects become important.

It is shown that the magic cut mainly scales with the hadronic centre-of-mass energy rather than with the mass of the $Z'$.

The validity of this picture relies heavily on large enough BR:
if additional decay modes of the $Z'$ are considered and the BR to charged leptons therefore reduced, the importance of interference increases.
$\Gamma / m$ is, in itself, \emph{not} a good measure of the size of interference effects.

Unless a full dedicated analysis is performed, in order to correctly interpret observed experimental limits in terms of specific models, it is important to compute the contribution of new physics in the \emph{same} kinematic range that is effectively being probed by the search.
Whether or not the model-dependent interference effects can then be safely neglected entirely depends on how large this kinematic range may be.
Within the classes of $Z'$ models considered here, using a $\Delta m / \sqrt{s}$ cut of no more than $\sim 5 \%$ ensures that the interference stays at the level of a few percent, which would thus allow an approximately model-independent analysis.\footnote{Here, ``model-independent'' essentially means that all the relevant features are captured by, for instance, the $c_u$-$c_d$ parameters, and that a more complete description such as given in table~\ref{tab:Zfactors} is superfluous.
It relies, however, on the absence of additional decay channels that would make the Z' wider.}
 For such  invariant mass window the relative contribution of the  interference 
 is comparable to the level of the combined  uncertainty 
from parton densities and higher order corrections~\cite{Accomando:2011eu}.

As closing statement, one can note that it is often important to use definitions of \BSM\ signal that include restrictions to a specific kinematic region not only from an experimental point of view (geometric acceptance, increased signal-over-background ratio) but also for theoretical reasons:
if some new physics is most visible within a given range, it does not automatically imply that the contribution in that region dominates the total cross-section.

\acknowledgments

EA, AB and SM are supported in part through the NExT Institute.
The CP$^3$-Origins centre is partially funded by the Danish National Research Foundation, grant number DNRF90.
DB would like to thank Laurent Thomas for useful discussions.

\bibliographystyle{JHEP}
\bibliography{references}

\end{document}